\documentclass[aps,pra,twocolumn,showpacs,letterpaper,tighten,float]{revtex4-1}
\usepackage{graphicx}
\usepackage{subfigure}
\usepackage{latexsym}
\usepackage{amsmath,amssymb}
\usepackage{bm} 
\usepackage{color}
\usepackage{stackrel}
\usepackage{epsfig}

\definecolor{myblue}{rgb}{.93, .93, 1}

\newcommand{\bsub}{\begin{subequations}}
	\newcommand{\esub}{\end{subequations}}

\begin{document}

\title{Localization-Driven Correlated States of Two Isolated Interacting Helical Edges}
\author{Yang-Zhi~Chou}\email{YangZhi.Chou@colorado.edu}

\affiliation{Department of Physics and Center for Theory of Quantum
  Matter, University of Colorado Boulder, Boulder, Colorado 80309,
  USA} \date{\today}
	
\begin{abstract}
We study the {\it localization-driven} correlated states among two isolated dirty interacting helical edges as realized at the boundaries of two-dimensional $\mathbb{Z}_2$ topological insulators. We show that an interplay of time-reversal symmetric disorder and strong interedge interactions generically drives the entire system to a gapless localized state, preempting all other intraedge instabilities. For weaker interactions, an antisymmetric interlocked fluid, causing a negative perfect drag, can emerge from dirty edges with different densities. We also find that the interlocked fluid states of helical edges are stable against the leading intraedge perturbation down to zero temperature.
The corresponding experimental signatures including zero-temperature and finite-temperature transport are discussed.
\end{abstract}

\maketitle

\section{Introduction}

Quenched randomness (disorder) can drastically suppress the electronic transport by inducing Anderson localization \cite{Anderson1958}, 
a phenomena that is known to be prominent in low dimensions.
Cooperations of interaction and disorder can induce manybody localization \cite{Basko06,Gornyi2005,Nandkishore2014} 
which exhibits ergodicity breaking and enables unexpected orders \cite{LPQO}.
As a striking outcome, a combination of time-reversal (TR) symmetric disorder and inter-particle interactions can drive a two-dimensional (2D) topological insulator \cite{Kane2005_1,Kane2005_2,Bernevig2006,Hasan2010_RMP,Qi2011_RMP,SenthilARCMP} (TI) edge, conducting ballistically in the absence of interaction \cite{Kane2005_1,Xie2016}, to a gapless insulating edge \cite{Chou2017}. 
In this work, we further explore the new correlated states due to a similar {\it localizing} mechanism among two isolated interacting $\mathbb{Z}_2$ TI edges with quenched disorder.

A 2D TR symmetric TI \cite{Kane2005_1,Kane2005_2,Bernevig2006,Hasan2010_RMP,Qi2011_RMP,SenthilARCMP} is a fully gapped bulk insulator whose edge is described by counter-propagating electrons forming Kramers pairs. 
The TR symmetry prevents the edge electrons from Anderson localization which generically ceases conductions in the conventional one-dimensional systems. 
Such a topological protected state emerges a helical Luttinger liquid description \cite{Wu2006,Xu2006} and exhibits a quantized $e^2/h$ edge conductance at zero temperature. 
The possibility of realizing 2D TR symmetric TI motivates various experimental studies \cite{Konig2007,Knez2011,Suzuki2013,Du2015,Li2015,Qu2015,Ma2015,Nichele2016,Nguyen2016,Couedo2016,Fei2017,Du2017,Li2017,Tang2017,Wu2018}
which might pave the way for creating Majorana and $\mathbb{Z}_4$ parafermion zero modes, enabling topological quantum computations \cite{Fu2009,Zhang2014,Orth2015,Alicea2016Review}.

\begin{figure}[t!]
	\includegraphics[width=0.35\textwidth]{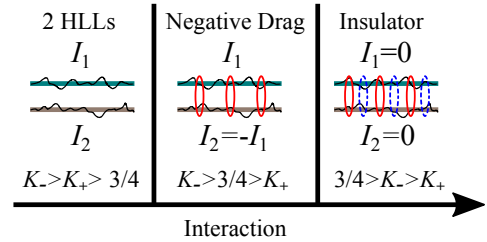}
	\caption{Zero-temperature phase diagram of two dirty TI edges with different densities. We assume $1>K_->K_+$ due to the repulsive interactions. For $K_->K_+>3/4$, the two helical Luttinger liquids are decoupled. The symmetric interedge mode is localized when $K_->3/4>K_+$. An anti-symmetric interlocked fluid is developed. For $3/4>K_->K_+$, the a gapless localized insulator is predicted. 
}
	\label{Fig:PD_neTI}
\end{figure}

Contrary to the well-studied single edge problems (see recent reviews \cite{Dolcetto2015,Rachel2018} and the references therein), the physics of two interacting TI edges \cite{Tanaka2009,Zyuzin10,Chou2015,Santos2015,Santos2016,Kainaris2017,Kagalovsky2018} has not been explored systematically, the effect due to simultaneous appearance of disorder and interactions especially.
In this work, we focus on the low temperature regimes of two isolated dirty interacting TI edges with different densities. 
We show that the combinations of interedge interactions and disorder can generate new types of localization-driven correlated states: A gapless insulating state with both edges being spontaneously TR symmetry broken, and an anti-symmetric interlocked fluid with edges carrying opposite currents. The former represents an interedge instability that preempts all other phases driven by TR symmetric intraedge perturbations. The latter corresponds to a zero temperature perfect {\it negative drag} in striking contrast with the well known perfect positive drag among quantum wires \cite{Nazarov1998,Klesse2000}.
These regimes are summarized in Fig.~\ref{Fig:PD_neTI}. We also discuss the stability of the negative drag state against intraedge perturbation.
Both of the interedge correlated states can be measured via a specific Coulomb drag \cite{Rojo1999,Narozhny2016} related experimental setup \cite{Chou2015} as illustrated in Fig.~\ref{Fig:drag} (a). 
Concomitantly, we predict the two terminal conductance at zero temperature (Fig.~\ref{Fig:cond}) and finite temperatures (Fig.~\ref{Fig:TD_cond}).

\section{Model}\label{Sec:Model}

We consider two isolated TR symmetric $\mathbb{Z}_2$ TI edges that interact via Coulomb force \cite{Tanaka2009,Zyuzin10,Chou2015,Kainaris2017} but do not allow interedge electron tunnelings. 
For each isolated edge, there are counter-propagating right ($R$) and left ($L$) mover fermions forming Kramers pairs.
In the low energy limit, the kinetic term is given by
\begin{align}\label{Free_F}
\hat{H}_0
=
-i \sum_{a=1,2} v_{Fa}
\int dx
\left[
R_a^{\dagger}(x) 
\partial_x 
R_a(x)
-
L_a^{\dagger}(x) 
\partial_x
L_a(x)
\right],
\end{align}
where $a=1,2$ is the edge index and $v_{Fa}$ is the Fermi velocity of the $a$th edge band. 
Time-reversal operation is encoded by $R_a(x)\rightarrow L_a(x)$, $L_a(x)\rightarrow -R_a(x)$, and $i\rightarrow -i$. 
Therefore, the conventional backscattering (e.g., $R^{\dagger}L+L^{\dagger}R$ in the spinless Luttinger liquid) is prohibited \cite{Kane2005_1}. 
The time-reversal symmetric disorder is the chemical potential fluctuation (pure forward scattering) given by,
\begin{align}
\hat{H}_V=\!\sum_{a=1,2}\int dx\,V_a(x)\!\left[R^{\dagger}_a(x)R_a(x)\!+\!L^{\dagger}_a(x)L_a(x)\right],
\end{align}
where $V_a(x)$ is the disordered potential in the $a$th edge. 
We assume that the disordered potentials are zero-mean Gaussian random variables and satisfy $\overline{V_a(x)V_{b}(y)}=\Delta\delta_{ab}\delta(x-y)$, where $\overline{\mathcal{O}}$ denotes a disorder average of $\mathcal{O}$.

The interaction between the two helical edges is primarily due to Coulomb interaction. Instead of studying specific microscopic models, we construct the interedge perturbations via symmetry and relevance in the renormalization group analysis. 
The leading TR symmetric backscattering terms are the interedge umklapp interactions \cite{Tanaka2009,Chou2015} given by
\begin{align}
\label{H_I_+}
\hat{H}_{U,+}=&\, U_+\int dx \left[e^{-i\delta Q_+x}L^{\dagger}_1R_1L^{\dagger}_2R_2	+ \text{H.c.}	\right],\\
\label{H_I_-}
\hat{H}_{U,-}=&\, U_-\int dx \left[e^{-i\delta Q_-x}L^{\dagger}_1R_1R^{\dagger}_2L_2	+ \text{H.c.}	\right].
\end{align}
In the above equations, $\delta Q_{\pm}=Q_{\pm}-2(k_{F1}\pm k_{F2})$ measures the lack of commensuration, $Q_{\pm}=2\pi/d$ is the commensuration wavevector ($d$ is the lattice constant of the 2D bulk), and $k_{F1}$ ($k_{F2}$) indicates the Fermi wavevector in the first (second) edge. 
Generically, both $\hat{H}_{U,-}$ and $\hat{H}_{U,+}$ are irrelevant due to incommensuration. 
We ignore the intraedge backscattering terms since they are subleading \cite{Schmidt2012,Kainaris2014,Chou2015}.

To include Luttinger liquid effects (arising from both intra- and interedge interactions), we use standard bosonization \cite{Giamarchi_Book,Shankar_Book}. 
The density ($n_a$) and current ($I_a$) can be expressed in terms of the phonon-like field ($\theta_a$). $n_a=\partial_x\theta_a/\pi$ and $I_a=-\partial_t\theta_a/\pi$.
The two helical Luttinger liquids problem can be decomposed into symmetric and anti-symmetric interedge degrees of freedom. 
In the imaginary time path integral, 
the bosonic action \cite{Klesse2000,Tanaka2009,Chou2015} is given by $\mathcal{S}_{\pm}=\mathcal{S}_{0,\pm}+\mathcal{S}_{V,\pm}+\mathcal{S}_{U,\pm}$, 
where
\begin{subequations}\label{Eq:S_pm}
	\begin{align}
	\mathcal{S}_{0,\pm}=&\frac{1}{2\pi v_{\pm} K_{\pm}}\int d\tau dx\,\left[\left(\partial_{\tau}\Theta_{\pm}\right)^2+v_{\pm}^2\left(\partial_x\Theta_{\pm}\right)^2\right],\\
	\label{Eq:S_Vpm}\mathcal{S}_{V,\pm}=&\int d\tau dx\,V_{\pm}(x)\frac{1}{\pi}\partial_x\Theta_{\pm},\\
	\label{Eq:S_Upm}\mathcal{S}_{U,\pm}=&\frac{U_{\pm}}{2\pi^2\alpha^2}\int d\tau dx\,\cos\left[2\sqrt{2}\Theta_{\pm}-\delta Q_{\pm}x\right],
	\end{align}
\end{subequations}
where $\Theta_{\pm}=\frac{1}{\sqrt{2}}\left[\theta_1\pm\theta_2\right]$ encodes the symmetric ($+$) and antisymmetric ($-$) collective modes, 
$K_{\pm}$ ($v_{\pm}$) is the Luttinger parameter (velocity), $V_{\pm}(x)=\frac{1}{\sqrt{2}}[V_1(x)\pm V_2(x)]$ is the disorder potential, and $\alpha$ is an ultraviolet length scale. 

The interedge Luttinger interaction is given by $(\partial_x\theta_1)(\partial_x\theta_2)\propto (\partial_x\Theta_+)(\partial_x\Theta_+)-(\partial_x\Theta_-)(\partial_x\Theta_-)$. 
As a consequence, repulsive interedge interactions
tend to decrease $K_+$ and increase $K_-$. 
[Note that $K_{\pm}<1$ ($K_{\pm}>1$) for overall repulsive (attractive) interactions.]
Importantly, the intraedge Luttinger interactions still dominate and drive $K_{\pm}<1$ \cite{Klesse2000}. 
We therefore assume that $1>K_->K_+$ holds generically.

Lastly, we discuss the disorder terms. $V_{\pm}(x)$ is a Gaussian random field which obeys $\overline{V_{\pm}(x)}=0$, $\overline{V_{\pm}(x)V_{\pm}(y)}=\Delta \delta(x-y)$, and $\overline{V_{+}(x)V_{-}(y)}=0$. The above conditions ensure that the symmetric and anti-symmetric sectors are completely decoupled. The intraedge perturbation will hybridize the two sectors. 
We will discuss the validity of our model in the end of the next section.

\section{Localization-Driven Correlated state}\label{Sec:LDCS}

We now discuss the zero temperature states in the simultaneously appearance of the interedge backscattering and the TR symmetric disorder. 
We will first review the mechanism that drives interedge collective modes into localization. Two new states (interedge localized and interlocked fluid states) can be inferred from the localization physics. We finally discuss the stability of the interlocked fluid states against intraedge perturbations.

\subsection{Interplay of disorder and interaction}

The two helical Lutinger liquids problem can be viewed as two decoupled problems of a disordered interacting helical edge \cite{Chou2017} with proper rescaling of parameters. We briefly review the ideas in Ref.~\onlinecite{Chou2017} and discuss the localization physics in this subsection.

We first discuss the stability of the Luttinger liquid phase.
The disorder potential $\mathcal{S}_{V,\pm}$ [given by Eq.~(\ref{Eq:S_Vpm})] generates chemical potential spatial fluctuations but does not induce backscattering. 
However, the interedge umklapp backscattering interaction $\mathcal{S}_{U,\pm}$ [given by Eq.~(\ref{Eq:S_Upm})] alone cannot gap out $\Theta_{\pm}$ unless $ |\delta Q_{\pm}|\le \delta Q_c$ \cite{PokrovskyTalapov} (where $\delta Q_c$ is the critical value in the commensurate-incommensurate transition). Therefore, the Luttinger liquid phase is generically stable with only disorder or interaction.
Nevertheless, the fluctuations of chemical potentials (equivalent to the fluctuations of $k_{F1}$ and $k_{F2}$) compensate the {\it missing} momenta ($\delta Q_{\pm}$) in a random fashion. As a result, the backscattering is enhanced due to ``local commensuration'' \cite{Fiete2006,Kainaris2014,Chou2015,Chou2017}.
Both the symmetric and anti-symmetric sectors in Eq.~(\ref{Eq:S_pm}) can be mapped to the localization problem studied in Ref.~\onlinecite{Chou2017} with a rescaling $K\rightarrow K_{\pm}/2$. 
The critical value $K_{\pm}=3/4$ \cite{CriticalK} (less interacting than the single edge critical value $K=3/8$ \cite{Wu2006,Xu2006,Chou2017}) separates a Luttinger liquid phase and a gapless localized phase.

For sufficiently strong interactions ($K_{\pm}<3/4$), the interedge $\Theta_{\pm}$ sector is driven to a localized state \cite{Giamarchi1988,Fisher1989,Chou2017} as the full gapped state (due to $\mathcal{S}_{U,\pm}$) is not stable against the {\it random field} disorder given by $\mathcal{S}_{V,\pm}$ \cite{Imry1975,Chou2017}.
In addition, the bosonized theory at $K_{\pm}=1/2$ can be mapped to a theory of massive Luther-Emery fermion with a chemical potential disorder \cite{Chou2017}, known to be Anderson localized for all the eigenstates \cite{Bocquet1999}. 
It can be further inferred that the physical state is a gapless insulator due to the structures of density and current operators in bosonization/refermionization \cite{Chou2017}.
Away from $K_{\pm}=1/2$, 
the refermionized theory becomes interacting and is no longer exactly solvable. 
For $K_{\pm}<1/2$, the backscattering is enhanced due to the additional repulsive interaction \cite{Kane1992PRB,Matveev1993,Garst2008} so the localization is stable. For $K_{\pm}>1/2$, the localization grows less stable as increasing $K_{\pm}$, and the critical point ($K_{\pm}=3/4$) is obtained from bosonization analysis.
The localizing mechanism here gives a nonmonotonic dependence in $\Delta$ with the strongest localization when $\Delta$ is comparable to $\delta Q_{\pm}$ \cite{Chou2017}.

\subsection{interedge localized state}

When both the symmetric and anti-symmetric sectors are localized ($K_+,K_-<3/4$), the edge state breaks TR symmetry spontaneously. 
We can define pseudospin operators for each edge \cite{Wu2006,Chou2017} whose finite expectation values indicate TR breaking of the localized states. 
The pseudospin expectation values in the localized state are random in space and uncorrelated among the two isolated edges.
The localized state here can be viewed two localized edges carrying half-charge \cite{Chou2017}. The Luther-Emery fermions at $K_+=K_-=1/2$ correspond to symmetric or the antisymmetric collective modes of the half-charge excitations among two edges.
Importantly, this interedge instability ($K_+,K_-<3/4$) dominates over the leading intraedge instability ($K<3/8$) \cite{Wu2006,Xu2006} because the critical interaction strength is weaker (larger Luttinger parameter).

\subsection{Interlocked fluid state}

For weaker interactions, there might exist a region such that only one of the interedge degrees of freedom is localized. The correlation among two edges is determined by the {\it remaining} delocalized collective mode. Such correlated states are called interlocked fluids in the studies of one-dimensional Coulomb drag and reflect the Luttinger liquid behavior \cite{Nazarov1998,Klesse2000,Laroche2014}. Here, we focus on the Coulomb drag physics among two generically unequal TI edges. This case was not considered in the existing literature.

For two isolated dirty TI edges with different electron densities, both the symmetric and anti-symmetric sectors are similar except $1<K_-<K_+$ (due to the repulsive interedge Luttinger interactions). 
A negative interlocked fluid can arise when $K_+<3/4$ and $K_->3/4$ since the symmetric sector is localized.
Such a correlated state is described by an anti-symmetric interedge collective mode, 
corresponding to a perfect ``negative drag''. 
In two dimensional electronic systems, a perfect negative drag can arise due to inter-layer exciton formation \cite{Su2008,Nandi2012}.
Similarly, a negative drag between two clean one-dimensional systems can also take place when the commensurate condition $|\delta Q_+|<\delta Q_c$ ($k_{F1}\approx-k_{F2}$) is finely tuned \cite{Chou2015,Furuya2015}.
Here, the interlocked anti-symmetric state is not induced by gapping at commensuration 
but by localizing the collective degrees of freedom. 
This localization-driven anti-symmetric interlocked fluid is also complementary to the early study for incommensurate clean quantum wires \cite{Fuchs2005}.
The phase diagram of the two dirty TI edges with different densities is summarized in Fig.~\ref{Fig:PD_neTI}.

As a comparison, for two clean TI edges with the same electron density ($k_{F1}=k_{F2}$), the interedge interaction $\mathcal{S}_{U,-}$ [given by Eq.~(\ref{Eq:S_Upm})] becomes to a commensurate backscattering term ($\delta Q_-=0$) that gaps out the anti-symmetric mode for $K_-<1$ \cite{Nazarov1998,Klesse2000} at zero temperature. 
The system therefore develops a symmetric interlocked fluid dictating a perfect positive drag \cite{Nazarov1998,Klesse2000,Chou2015}. In the presence of disorder, the symmetric interlocked fluid remains stable as long as $K_+>3/4$. 
The fully gapped anti-symmetric mode becomes to a gapless localized state because the long range order is unstable against {\it random field} disorder in one dimension \cite{Imry1975,Chou2017}. 
For $K_+<3/4$, the system develops an interedge fully localized state that halts conduction at all.

\begin{figure}[t!]
\includegraphics[width=0.4\textwidth]{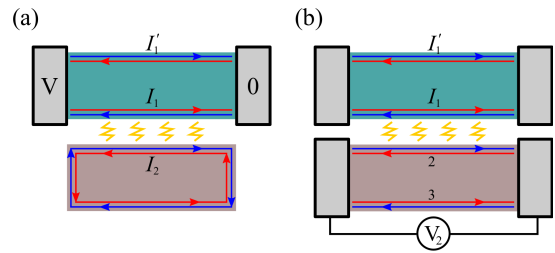}
\caption{(a) The proposed experimental setup \cite{Chou2015} (``edge gear'') for studying the interedge correlated states. The top TI is attached to two external electrodes which result in two separated edges carrying current $I_1$ and $I_1'$; the bottom TI forms a close edge loop with a current $I_2$ (but without a voltage drop). The two proximate edge states (carrying currents $I_1$ and $I_2$) interact via interedge Coulomb interactions. As discussed in the main text, the two terminal conductance on the top TI encodes the information of the interedge correlated states. 
(b) The standard Coulomb drag experiment in the lateral geometry as a comparison. 
}
\label{Fig:drag}
\end{figure}

\subsection{Stability of interlocked fluid states}\label{Sec:stability}

In Ref.~\onlinecite{Ponomarenko2000}, the stability of a perfect drag against the single-particle impurity scattering was investigated. The impurity scattering with in a quantum wire can hybridize the symmetric and anti-symmetric collective modes. As a consequence, the perfect drag is only stable above certain temperature scale set by disorder scattering \cite{Ponomarenko2000}. Here, we repeat the same analysis for drags among two helical Luttinger liquids.

Due to the TR symmetry, the single-particle backscattering (e.g., $L^{\dagger}_1R_1$) is not allowed. Therefore, we consider the TR symmetric impurity two-particle backscattering interaction \cite{Wu2006,Maciejko2009} as follows:
\begin{align}
\hat{H}_{\text{imp}}=\sum_{a=1,2}W_a \left[
L_a^{\dagger}(0+\alpha)L_a^{\dagger}(0)R_a(0)R_a(0+\alpha)
+\text{H.c.}  \right],
\end{align}
where $W$ is the strength of impurity interaction and a point splitting with the ultraviolet length $\alpha$ is performed. The corresponding bosonic action is
\begin{align}
\nonumber\mathcal{S}_W=&\sum_{a=1,2}\tilde{W_a}\int d\tau
\,\cos\left[4\theta_a(\tau,x=0)\right]\\
\nonumber=&\tilde{W_1}\int d\tau
\,\cos\left[2\sqrt{2}\left(\Theta_++\Theta_-\right)\right]\\
\label{Eq:H_W}&+\tilde{W_2}\int d\tau
\,\cos\left[2\sqrt{2}\left(\Theta_+-\Theta_-\right)\right]
\end{align}
where $\tilde{W}_a=W_a/(2\pi^2\alpha^2)$. Based on the scaling dimensions \cite{Wu2006,Maciejko2009}, $\tilde{W}_1$ and $\tilde{W}_2$ become relevant when $K_++K_-<1/2$. These intraedge interactions are the sub-leading perturbations because the interedge localizaiton happens when $K_{\pm}<3/4$. (As a comparison, the clean helical Luttinger liquid drag happens when $K_-<1$ \cite{Chou2015}.)

To further investigate the stability of the interlocked fluid states, we follow the treatment in Ref.~\onlinecite{Ponomarenko2000}. We focus on the antisymmetric interlocked fluid (negative drag) for $K_+<3/4$ and $K_->3/4$. Then, we assume the symmetric sector is in the semicalssical limit ($K_+\rightarrow 0^+$). In such an approximation, the $\Theta_+(\tau,x)$ can be replaced by a time-independent function $\gamma_+(x)$, and all the contributions from instanton tunnelings between degenerate vacuums are ignored. Enabling the instanton tunneling will make the impurity scattering less relevant, so the semiclassical treatment here can be viewed as ``the worst case scenario.''
The impurity two-particle backscattering interaction is approximated by $\cos(2\sqrt{2}\Theta_-+C)$, where $C$ is an unimportant constant. As a consequence, Eq.~(\ref{Eq:H_W}) becomes relevant when $K_-<1/2$. This analysis confirms that the antisymmetric interlocked fluid ($K_+<3/4<K_-$) remains stable when the symmetric mode is fully localized. The same stability also applies to the symmetric interlocked fluid due to two helical liquids with the same density for $K_+>1/2$. 

In conclusion, the intraedge perturbations do not sabotage the interlocked fluid states among two helical Luttinger liquid, in contrast to the conventional Coulomb drag \cite{Nazarov1998,Klesse2000} where the stability against the impurity backscattering is only valid for temperatures higher than the scale set by disorder \cite{Ponomarenko2000}. The stability of drag among helical liquids is a manifestation of the topological protection in the topological insulator edges.

\section{Proposed experimental setup}\label{Sec:Exp}

The physics of two isolated TI edges is related to the Coulomb drag experiments \cite{Yamamoto2006,Laroche2014,Rojo1999,Narozhny2016} in one dimensional systems. We focus on the ``edge gear'' setup \cite{Chou2015} [in Fig.~\ref{Fig:drag} (a)] that detects all the interedge correlated states discussed above. 
We will first focus on infinitely long edges at zero temperature. The corrections due to finite sizes and/or finite temperatures are discussed via existing well-known properties of the localized insulator and Luttinger liquid analysis.

\subsection{Edge gear setup: Results with an infinite-long size at zero temperature}

The edge gear setup \cite{Chou2015} in Fig.~\ref{Fig:drag} (a) contains two isolated TI systems in the lateral geometry. Two TIs are separated via a gap such that two proximate edges can interact via Coulomb force, but the electron tunneling is prohibited.
The top TI is connected to two external leads while the bottom TI forms a close edge loop.
The two terminal conductance is measured in the top TI system whose value generically encodes the interedge correlation.


\begin{figure}[t!]
	\includegraphics[width=0.3\textwidth]{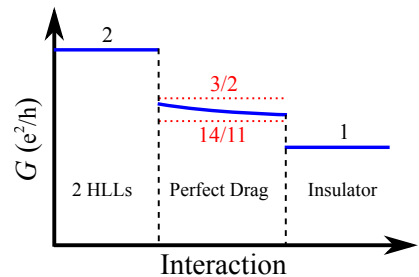}
	\caption{The two terminal conductance at zero temperature as a function of interaction in the edge gear setup [Fig.~\ref{Fig:drag} (a)]. 
		For sufficiently weak interedge interactions, the conductance (in the unit of $e^2/h$) is $2$ as the absence of the bottom close loop TI. In the perfect drag regime, the conductance follows Eq.~(\ref{Eq:G_gear}) with an upper bound $3/2$ and a lower bound $14/11$ (red dotted lines). These bounds guarantee discontinuities of the conductance. 
		For sufficiently strong interactions, two TI edge states become localized insulators. The conductance becomes to 1.
	}
	\label{Fig:cond}
\end{figure}


Firstly, in the absence of any interedge interaction, the conductance is $2\frac{e^2}{h}$ (due to two edge channels) independent of the Luttinger parameter \cite{Safi1995,Maslov1995,Ponomarenko1995}. 
For both $K_+,K_-<3/4$, the interedge localized state takes place and makes $I_1=I_2=0$. 
The conductance is therefore reduced to $\frac{e^2}{h}$ as only the edge with current $I'_1$ is conducting. 
For the interlocked fluids, 
the interedge interactions induce $I_1=\pm I_2$ where the positive or negative sign corresponds to the perfect positive or negative drag. The conductance (for both the positive and negative drags) \cite{Chou2015} is
\begin{align}\label{Eq:G_gear}
G=\frac{I_1'+I_1}{V}=\frac{e^2}{h}\left[1+\frac{1}{1+1/K}\right]
\end{align}
which encodes the Luttinger parameter $K$ \cite{LuttingerK} of the close loop TI edge state.
The nonuniversal conductance varies from $\frac{3}{2}\frac{e^2}{h}$ ($K=1$, noninteracting limit) to $\frac{14}{11}\frac{e^2}{h}$ ($K=3/8$, intraedge instability \cite{Xu2006,Wu2006}). 
As plotted in Fig.~\ref{Fig:cond}, those bounds ensure two stage conductance ``transitions'' (discontinuities) when tuning the interaction. 
We note that Eq.~(\ref{Eq:G_gear}) is based on the ``Luttinger liquid lead'' approximation \cite{Chou2015}. For an ideal close loop (infinite coherence time) in the perfect drag regime \cite{Horovitz2018}, the conductance is predicted to be $2e^2/h$ as if the close loop was absent.

The only missing ingredient from the edge gear setup is the ``sign'' (positive/negative) of the perfect drag since the two terminal conductance in Eq.~(\ref{Eq:G_gear}) only encodes the electron correlation. 
A separate measurement (e.g., imaging edge currents via SQUID \cite{Nowack2013,Spanton2014}) is required for revealing the parallel and antiparallel nature of the interlocked fluid states.


\begin{figure}[t!]
	\includegraphics[width=0.4\textwidth]{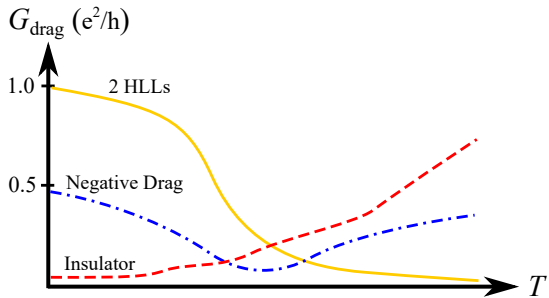}
	\caption{The sketched temperature dependence of drag conductance in various regimes. $G_{\text{drag}}=G-\frac{e^2}{h}$ where $G$ is the two-terminal conductance of edge gear setup. The yellow solid line indicates the two helical liquids regime ($K_+,K_->3/4$); the blue dot-dashed line indicates the negative drag ($K_+<3/4<K_-$); the red dashed line indicates the localized regime ($K_+,K_-<3/4$).  
	The detail features of each curve are explained qualitatively in the main text. 
	}
	\label{Fig:TD_cond}
\end{figure}


\subsection{Edge gear setup: Finite-size and finite-temperature corrections}

Now, we discuss the finite size and the finite temperature corrections. 
All the localization-driven correlated states predicted in this work require the edge length $L\gg \xi_{\text{loc}}$ where $\xi_{\text{loc}}$ is the localization length.
The drag conductance ($G_{\text{drag}}=I_1/V$) can be expressed by $G_{\text{drag}}=G_++G_-$, where $G_+$ and $G_-$ are the conductance contributions due to the symmetric and antisymmetric sectors, respectively. 

For delocalized modes ($K_{\pm}>3/4$), the primary sources of perturbations come from the inelastic scattering due to $\hat{H}_{U,\pm}$. The leading conductance correction is given by $\delta G_{\pm}=G_{\pm}-G_{\pm,0}\propto-T^{4K_{\pm}-2}$ for $T\ll \Delta/v$ \cite{Chou2015}, where $G_{\pm,0}$ is the conductance at zero temperature. At sufficiently high temperature, we can deduce the conductance via the conductivity of the Luttinger liquid analysis \cite{Chou2015}.
For $T\gg v|\delta Q_{\pm}|$, the conductance is given by $G_{\pm}=\frac{\sigma_{\pm}}{L}\propto T^{-4K_{\pm}+3}$ \cite{Chou2015}, where $\sigma_{\pm}$ is the conductivity of the symmetric/anti-symmetric sector.

For $K_{+}<3/4$ ($K_-<3/4$), the symmetric (anti-symmetric) mode becomes localized. 
In a finite length localized insulator, there exist multiple temperature regimes \cite{Imry2002}. For sufficiently high temperatures, the thermal length is smaller than the localization length so the Luttinger liquid analysis can be applied \cite{Nazarov1998,Klesse2000,Fiete2006,Chou2015}. 
We summarize the temperature dependence as follows:
\begin{align}\label{Eq:loc_TD}
G_{\pm}^{\text{loc}}\propto\begin{cases}
e^{-2L/\xi_{\text{loc},\pm}},& \text{for }T\ll T'_{\pm}\\[2mm]
e^{-\text{const}\sqrt{T_{0,\pm}/T}},& \text{for }T'_{\pm}\ll T\ll T_{0,\pm}\\[2mm]
e^{-T_{0,\pm}/T},& \text{for }T_{0,\pm}\ll T<\delta E_{m,\pm},\\[2mm]
T^{-4K_{\pm}+2}, & \text{for }\delta E_{m,\pm}\ll T\ll \Delta/v\\[2mm]
T^{-4K_{\pm}+3}, & \text{for }T\gg \delta E_{m,\pm}, v|\delta Q_{\pm}|
\end{cases}
\end{align}
where $\xi_{\text{loc},\pm}$ is the localization length in the symmetric or antisymmetric sector, $T'_{\pm}$ and $T_{0,\pm}$ correspond to the lower and upper bounds of the variable range hopping mechanism \cite{Mott1966electrical,Imry2002}, and $\delta E_{m,\pm}$ indicates the distance between mobility edge energy and the fermi energy in a finite-size 1D insulator. $T'_{\pm}\equiv v\xi_{\text{loc},\pm}/L^2$ is determined by setting the optimal hopping length to be the same as the finite edge length $L$; $T_{0,\pm}\equiv v/\xi_{\text{loc},\pm}$ corresponds to the typical energy separation in a localized length $\xi_{\text{loc},\pm}$.
For $T\gg \delta E_{m,\pm}$, the localized state is no longer sharply defined. We can treat the backscattering interactions as perturbations with the Luttinger liquid analysis. The standard drag conductivity predicts two high temperature regimes \cite{Chou2015,Fiete2006} similar to the results for $K_{\pm}>3/4$. The regime yields $T^{-4K_{\pm}+2}$ will disappear if $\delta E_{m,\pm}\ge \Delta/v$. We note that the conductance $G_{\pm}^{\text{loc}}$ is at most $\frac{1}{2}e^2/h$. 

Combining the results above, we summarize the temperature dependence in the three regimes. 
All the results are summarized in Fig.~\ref{Fig:TD_cond}.
In the decoupled helical liquids regime ($K_{+},K_->3/4$), the measured two-terminal conductance ($G_L$) is given by \cite{Chou2015}
\begin{align}
G_L\!(T)=\!\begin{cases}
2\frac{e^2}{h}-A_1T^{4K_{+}-2}-A_2T^{4K_--2},& \text{for }T\ll \Delta/v,\\[2mm]
\frac{e^2}{h}+B_1T^{-4K_++3}+B_2T^{-4K_-+3},& \text{for }T\gg v|\delta Q_{\pm}|,
\end{cases}
\end{align}
where $A_{1,2}$ and $B_{1,2}$ are temperature-independent constants. 
The conductance is monotonically decreasing as increasing $T$ in this regime. The size dependence is absorbed into $A_{1,2}$ and $B_{1,2}$

In the interedge localized regime ($K_+,K_-<3/4$), the conductance is $G_L(T)=\frac{e^2}{h}+G_+^{\text{loc}}+G_-^{\text{loc}}$ where $G_{+}^{\text{loc}}$ and $G_{-}^{\text{loc}}$ are given by Eq.~(\ref{Eq:loc_TD}). The highest temperature regime gives a temperature enhancing conductance behavior because $-4K_{\pm}+3>0$. 
The conductance is essentially a monotonically increasing function of temperature. The potential nonmonotonicity is in the vicinity of  $T\sim \Delta/v$ when $1/2<K_{\pm}<3/4$.

In the negative drag regime $(K_+<3/4,K_->3/4)$, the zero temperature conductance of a finite size system is $G_L(0)=G+C\frac{e^2}{h}e^{2L/\xi_{\text{loc},+}}$ where $G$ is given by Eq.~(\ref{Eq:G_gear}) and $C$ is a constant. The temperature dependent conductance is given by
\begin{align}
G_L(T)\!=\!\begin{cases}
G_L(0)-D_1 T^{4K_--2},&  \text{for }T\!\ll \Delta/v,\\[2mm]
\frac{e^2}{h}+\frac{D_2}{T^{4K_+-3}}+\frac{D_3}{T^{4K_--3}},& \text{for }\!T\gg v|\delta Q_{+}|,\delta E_m,
\end{cases}
\end{align}
where $D_1$, $D_2$, and $D_3$ are constants. At high temperatures, the $D_2$ term wins over $D_3$ term because $4K_+-3<0<4K_--3$. 
The conductance in the negative drag regime is a non-monotonic function in temperature. The non-monotonicity can be understood by the interplay of the localized symmetric mode (monotonically increasing conductance) and delocalized antisymmetric mode (monotonically decreasing conductance).

\subsection{Drag resistivity setup}

As a comparison, we discuss the standard ``drag resistivity'' setup \cite{Rojo1999,Narozhny2016} as illustrated in Fig.~\ref{Fig:drag} (b). The drag resistance is defined by $R_D=-V_2/I_1$. $V_2$ is the generated voltage canceling the electromotive force due to the interedge interaction. Both the interedge localized and the interlocked fluid states tend to develop infinite zero-temperature drag resistivity $\rho_D=R_D/L$ (where $L$ is the length of edge). 
The sign of the perfect drag can be measured in principle. 
Meanwhile, the interedge localized state also contributes a nonuniversal sign which is determined by the {\it weaker} localized interedge collective mode.
We therefore conclude that there is no simple way to separate interedge localized and interlocked fluid states 
from the standard setup in the zero-temperature limit. 
In addition to the above mentioned issues, the edge 3 in the bottom TI [of Fig.~\ref{Fig:drag} (b)] most likely shorts the system.

\section{Summary and discussion}\label{Sec:summary}

We have studied the zero temperature phases in two isolated dirty interacting TI edges. We showed that an interedge localized state can generically takes place due to an interplay of TR symmetric disorder and interedge interactions. 
We also predicted that an anti-symmetric interlocked fluid state, producing a negative drag, can arise among two dirty TI edges with different densities. The anti-symmetric interlocked fluid is a consequence of localized symmetric collective mode and delocalized antisymmetric collective mode. 
Moreover, the interlocked fluids states among two TI edges is founded to be stable down to zero temperature, in contrast to the quantum wire systems where the drag is only valid above some temperature corresponding to disorder scattering \cite{Ponomarenko2000}. 
Our study explicitly shows that non-trivial interedge correlations can still arise even without commensuration.
The zero- and finite-temperature transport signatures of the edge gear setup \cite{Chou2015} are discussed.

We comment on the negative drag between two generically unequal TI edges. 
This scenario is specific to TI edge states where single particle backscattering is absent, so the negative drag can be viewed as a signature of Coulomb drag among helical Luttinger liquids. 
The condition of different densities is reminiscent of the experimental observation of negative drag among asymmetric quantum wires \cite{Yamamoto2006} whose mechanism has not been concluded yet. 
Our results might provide a new perspective for understanding the negative drag in one dimensional systems.

In this work, we merely consider sufficiently long TI edges within the standard Luttinger liquid analysis and the linear response theory. The effect of dispersion nonlinearity \cite{Imambekov2012RMP} and the finite electric field response \cite{Nattermann2003} are interesting future directions.
The finite close edge loop correction in the edge gear setup [Fig.~\ref{Fig:drag} (a)], potentially generating a resonant feedback for an ac drive, is an interesting topic in the future.

{\it Acknowledgment.}-- We thank Matthew Foster, Chang-Tse Hsieh, Tingxin Li, Rahul Nandkishore, Leo Radzihovsky, and Zhentao Wang for useful discussions.
We are also grateful to Rahul Nandkishore for the useful feedback on this manuscript.
This work is supported in part by a Simons Investigator award to Leo Radzihovsky and in part by the Army Research Office under Grant Number W911NF-17-1-0482.
The views and conclusions contained in this document are those of the authors and should not be interpreted as representing the official policies, either expressed or implied, of the Army Research Office or the U.S. Government. The U.S. Government is authorized to reproduce and distribute reprints for Government purposes notwithstanding any copyright notation herein.


%


\end{document}